\documentclass[sigconf]{acmart}
\usepackage{enumitem}
\usepackage{pifont}
\usepackage{color, colortbl}
\definecolor{LightGray}{rgb}{0.92,0.92,0.92}
\usepackage{multirow}
\usepackage[noend]{algpseudocode}
\usepackage{algorithmicx,algorithm}
\copyrightyear{2023}
\acmYear{2023}
\setcopyright{acmlicensed}
\acmConference[SIGIR '23] {Proceedings of the 46th International ACM SIGIR Conference on Research and Development in Information Retrieval}{July 23--27, 2023}{Taipei, Taiwan.}
\acmBooktitle{Proceedings of the 46th International ACM SIGIR Conference on Research and Development in Information Retrieval (SIGIR '23), July 23--27, 2023, Taipei, Taiwan}
\acmPrice{15.00}
\acmISBN{978-1-4503-9408-6/23/07}
\acmDOI{10.1145/3539618.3591705}

\begin{document}

\title{
Keyword-Based Diverse
Image Retrieval by\\Semantics-aware Contrastive Learning and Transformer}

\author{Minyi Zhao}
\authornote{Major part of this work was done while the author was an intern at Tencent.}
\email{zhaomy20@fudan.edu.cn}
\affiliation{%
  \institution{School of Computer Science\\Fudan University}
  \city{Shanghai}
  \country{China}
}

\author{Jinpeng Wang}
\email{wjp20@mails.tsinghua.edu.cn}
\affiliation{%
  \institution{Tsinghua Shenzhen Int'l Graduate School, Tsinghua University}
  \city{Shenzhen}
  \country{China}
}

\author{Dongliang Liao}
\email{brightliao@tencent.com}
\authornote{Corresponding author.}
\affiliation{%
  \institution{Wechat Group\\Tencent Inc.}
  \city{Guangzhou}
  \country{China}
}

\author{Yiru Wang}
\email{dorisyrwang@tencent.com}
\affiliation{%
  \institution{Wechat Group\\Tencent Inc.}
  \city{Beijing}
  \country{China}
}

\author{Huanzhong Duan}
\email{boosterduan@tencent.com}
\affiliation{%
  \institution{Wechat Group\\Tencent Inc.}
  \city{Beijing}
  \country{China}
}

\author{Shuigeng Zhou}
\authornotemark[2]
\email{sgzhou@fudan.edu.cn}
\affiliation{%
  \institution{School of Computer Science\\Fudan University}
  \city{Shanghai}
  \country{China}
}
\renewcommand{\shortauthors}{Minyi Zhao et al.}
\begin{abstract}
In addition to relevance, diversity is an important yet less studied performance metric of cross-modal image retrieval systems, which is critical to user experience.
Existing solutions for diversity-aware image retrieval either explicitly post-process the raw retrieval results from standard retrieval systems or try to learn multi-vector representations of images to represent their diverse semantics. However, neither of them is good enough to balance relevance and diversity.
On the one hand, standard retrieval systems are usually biased to common semantics and seldom exploit diversity-aware regularization in training, which makes it difficult to promote diversity by post-processing.
On the other hand, multi-vector representation methods are not guaranteed to learn robust multiple projections. As a result, irrelevant images and images of rare or unique semantics may be projected inappropriately, which degrades the relevance and diversity of the results generated by some typical algorithms like top-$k$. To cope with these problems, this paper presents a new method called CoLT that tries to generate much more representative and robust representations for accurately classifying images.
Specifically, CoLT first extracts semantics-aware image features by enhancing the preliminary representations of an existing one-to-one cross-modal system with semantics-aware contrastive learning. Then, a  transformer-based token classifier is developed to subsume all the features into their corresponding categories. Finally, a post-processing algorithm is designed to retrieve images from each category to form the final retrieval result. Extensive experiments on two real-world datasets Div400 and Div150Cred show that CoLT can effectively boost diversity, and outperforms the existing methods as a whole (with a higher $F1$ score).
\end{abstract}

\ccsdesc[500]{Information systems~Information retrieval diversity}

\keywords{Cross-modal retrieval, Keyword-based image
retrieval, Diversification retrieval, Transformer}

\maketitle

\section{Introduction}
\begin{figure}[t]
    \centering
    \includegraphics[width=0.9\columnwidth]{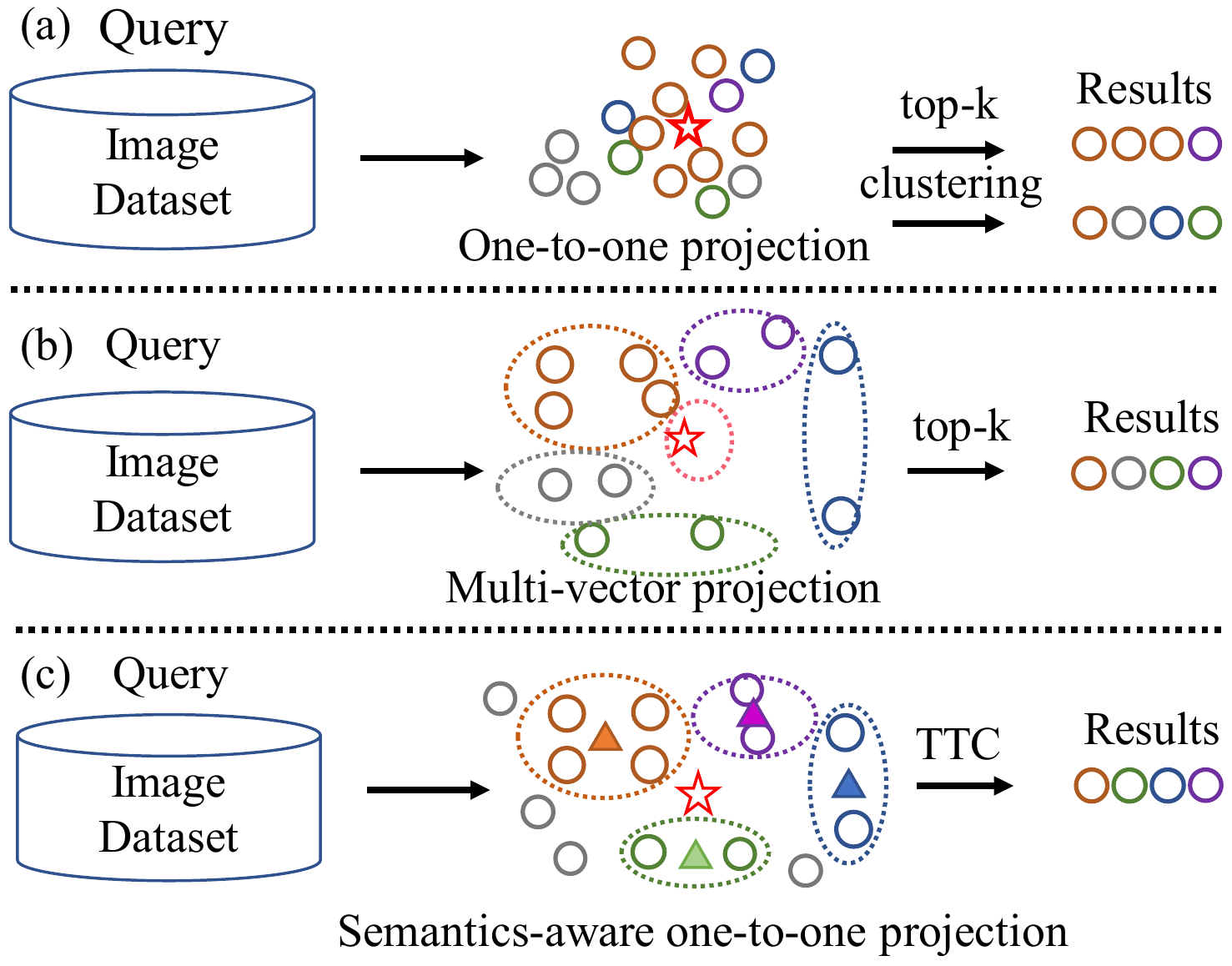}
    \caption{Illustrations of (a) typical cross-modal image retrieval systems; (b) learning-based multi-vector retrieval systems; (c) our CoLT method. Red star represents the query. Points of different colors denote images of different semantics. Gray points represent irrelevant images, and triangles represent the prototypes of the corresponding semantics. Dotted circles denote the projection regions.}
    \label{fig:motivation}
\end{figure}

With the popularity of the Web and its applications, increasing data are created and posted to the Web, which triggers the rapid development of web search and information retrieval techniques~\cite{singhal2001modern,yang2021limited}. Among them, cross-modal data retrieval~\cite{wang2016comprehensive,wang2017adversarial,zhen2019deep,cao2020adversarial,zeng2021multi,wang2017adversarial,cai2021neural,zeng2022moment} enables users to conveniently acquire desirable information in different forms. A typical example is cross-modal image retrieval (CMIR in short)~\cite{wang2019camp,wang2016learning,chen2020imram,rasiwasia2010new,zhai2014modeling,young2014from}. CMIR takes a textual query as input to retrieve images with matched semantics, has been deployed in many web applications like Instagram and Flickr, and gains increasing research attention~\cite{gu2021deep,liu2021social,ji2021attribute,xu2017semisuper,liao2019popularity}.

Nowadays, the relevance of cross-modal image retrieval systems has been significantly advanced by recent works~\cite{wang2019camp,liu2021inflate,liu2021image,yu2022u} and large-scale one-to-one pre-trained encoders~\cite{radford2021learning,xu2022groupvit} with the help of massive image-text pairs crawled from the web. However, existing models are prone to return a list of retrieved images with similar semantics. Often, the queries submitted by ordinary users, especially those in the form of short keyword-based texts without concrete context~\cite{qin2020diversifying,wang22spatiotemporal}, very likely have broad semantics, and thus are semantically ambiguous and uncertain. For example, given the coarse-grained query ``dog", the user may expect dog images with diverse semantics (\textit{e.g.} different breeds, colors, and body shapes). Obviously, keyword-based queries are prone to match various retrieval results, but a list of images with similar semantics cannot meet the diverse requirements of different users, thus deteriorating their retrieval experience~\cite{su2021modeling,zeng2022keyword}.

To address the aforementioned drawback, the task of \emph{keyword-based diverse image retrieval}~\cite{inoescu2016retrieving,inoescu2014div400,ionescu2016result,ionescu2020benchmarking}, is proposed, which takes a short keyword-based text as input to search a list of images with high relevance and rich semantic diversity. Recent approaches can be roughly divided into two groups. The first group is post-processing based approaches~\cite{seddati2017umons,peng2017cfm,zaharieva2014auf,renders2017nle,seddati2017umons,sarac2014retina,gamzu2020query,kuzi2019query}. These methods usually apply existing cross-modal encoders to extracting features. Then, various algorithms (\textit{e.g.} re-ranking~\cite{zaharieva2014auf} and clustering~\cite{peng2017cfm}) are adopted to promote the diversity. However, these methods often cannot obtain a good retrieval list with balanced relevance and diversity, due to the limitations of \emph{one-to-one projection}. For instance, as shown in Fig.~\ref{fig:motivation}(a), on the one hand, in typical one-to-one projection, (\textbf{W1}) the
query feature (\emph{the red star}) is likely to be surrounded by images of common semantics (\emph{the brown points}) due to the long-tail distribution of the training data, which will make the top-$k$ result set full of images of similar semantics. On the other hand, (\textbf{W2}) image features with different semantics are less distinguishable because of the ignorance of modeling diversity~\cite{zeng2022keyword}, which will hurt the performance of some algorithms like clustering.

The second group is a set of learning-based approaches~\cite{bo2019diversified,zhao2017deep,su2021modeling,song2019polysemous,wu2020interpretable,zeng2022keyword} that try to use various techniques (\textit{e.g.} graph~\cite{su2021modeling}, metric learning~\cite{chen2020imram,wang2017adversarial} and multiple instance learning~\cite{zhao2017deep,xu2020proposal}) to model the diversity. Compared with the one-to-one projection that projects each image to a vector in the latent space, these methods~\cite{wu2020interpretable,song2019polysemous,zeng2022keyword} embed each image (or text query) into multiple vectors around the relevant features to obtain their diverse representations for top-$k$ search, namely \emph{multi-vector projection}. Unfortunately, such a projection is not robust enough and unable to handle images of rare or unique semantics. As shown in Fig.~\ref{fig:motivation}(b), (\textbf{W3}) some irrelevant outliers (\emph{the grey points}) will be mistakenly projected to represent diversity. Besides, (\textbf{W4}) some images of rare or unique semantics (\emph{the blue points}), will very possibly be projected into some remote regions where the top-$k$ algorithm cannot reach.

To overcome the weaknesses (\textit{i.e.,} \textbf{W1}$\sim$\textbf{W4}) of the existing methods, in this paper we propose a novel approach called CoLT (the abbreviation of Semantics-aware \textbf{Co}ntrastive \textbf{L}earning and \textbf{T}ransformer) for keyword-based image retrieval. In particular, to overcome \textbf{W1}, \textbf{W2} and \textbf{W3}, CoLT extracts stable, representative and distinguishable image features with the help of a new \emph{semantics-aware contrastive learning} (SCL) loss. As shown in Fig.~\ref{fig:motivation}(c), the core idea of SCL is to project images of similar semantics (\textit{e.g.} dogs of the same breed) to vectors around their matched semantic prototype that keeps a proper distance from the other prototypes (\textit{e.g.} dogs of different breeds and irrelevant images) and the query feature to better model the diversity. As for coping with images of rare semantics (\textbf{W4}), instead of utilizing top-$k$ algorithm as in existing works, CoLT employs a powerful \emph{transformer-based token classifier} (TTC) to generate the final retrieval results. Specifically, in TTC the image and query features are concatenated as an input token sequence. Subsequently, TTC classifies each token into a relevant semantic category to distinguish the images of various semantics. Finally, a flexible post-processing algorithm is designed to select images from various semantic categories (both common and rare semantics), to form the final retrieval results.
Such a design offers our method four-fold advantages: (i) \textit{High semantic relevance.} CoLT improves the robust one-to-one projection of pre-trained cross-modal encoders, which is much more stable than recent multi-vector projection-based methods. (ii) \textit{High semantic diversity.} CoLT not only makes the image features much more distinguishable via semantics-aware contrastive learning but also uses a transformer-based token classifier to mine rare semantics. (iii) \textit{General and easy-to-use.} CoLT can be directly stacked at the end of various existing cross-modal encoders, without modifying their structures and parameters, and boost the performance in a plug-and-play manner. (iv) \textit{Easy-to-control.} We can modify the post-processing algorithm in CoLT to flexibly balance semantic relevance and semantic diversity without re-implementing the model.

Contributions of this paper are summarized as follows: (1) We pinpoint the limitations of existing methods and present a novel approach called CoLT for keyword-based diverse image retrieval. CoLT first extracts high-quality and distinguishable semantics-aware features and then classifies the features to generate the final retrieval list. (2) We develop a semantics-aware contrastive loss in CoLT to extract more robust and representative features. (3) To better mine semantic diversity, we design a transformer-based token classifier to generate the retrieval results. (4) We conduct extensive experiments on two real-world datasets Div400 and Div150Cred, which show that our method can effectively boost the diversity, and outperforms the existing methods as a whole with a higher $F1$ score.

\section{Related Work}
\subsection{Cross-Modal Image Retrieval}
Typical cross-modal image retrieval methods~\cite{peng2018overview} can be roughly divided into two categories: cross-modal similarity measurement based methods~\cite{tong2005graph,zhuang2008mining,yang2011multimedia,zhai2012effective} that directly calculate the cross-modal distance and common space learning-based methods~\cite{zhen2019deep,wang2017adversarial,chen2019cross,lou2017simple,yan2020semantics} that map the query and images into a shared space via various techniques like attention mechanism and generative adversarial network etc. Nowadays, thanks to the transformer structure and pre-training techniques, large-scale pre-trained encoders (\textit{e.g.} CLIP~\cite{radford2021learning}, ALIGN~\cite{jia2021scaling}, GroupViT~\cite{xu2022groupvit}, and U-BERT~\cite{yu2022u}) have shown their superiority in relevance-based retrieval tasks. Although these methods have significantly improved the retrieval relevance, their ignorance of modeling semantic diversity hurts the semantic diversity of their retrieval lists.
\begin{figure*}[t]
    \centering
    \includegraphics[width=2.0\columnwidth]{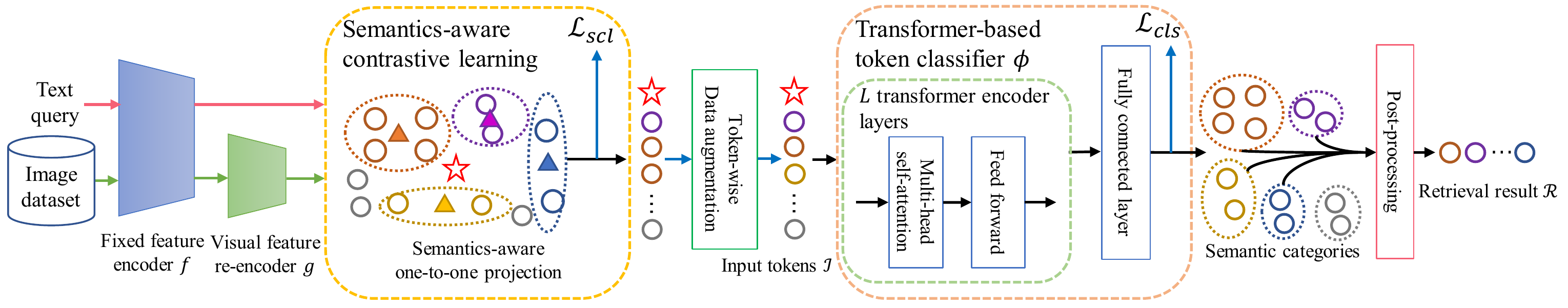}
    \caption{The architecture of CoLT. Blue lines are valid only during training.}
    \label{fig:pipeline}
\end{figure*}

\subsection{Diverse Retrieval}
Existing diverse retrieval approaches roughly fall into two groups. The first group is post-processing based methods~\cite{seddati2017umons,peng2017cfm,zaharieva2014auf,renders2017nle,seddati2017umons,sarac2014retina,gamzu2020query,kuzi2019query,chen2018fast}, which usually use existing feature encoders~\cite{le2014distributed,he2016deep,radford2021learning,devlin2019bert,dosovitskiy2020image} to generate features, then mine the diversity with a post-processing algorithm. Among them, \cite{seddati2017umons} first filters irrelevant images, then clusters the rest via DBSCAN~\cite{ester1996density} to promote diversity. MMR~\cite{renders2017nle} is proposed to re-rank the retrieval list to balance diversity and relevance. \citet{bo2019diversified} extract keywords to control the diversity of the results.
The second group includes recently proposed learning-based methods, which aim to represent the semantic diversity in the latent space~\cite{bo2019diversified,zhao2017deep,su2021modeling,song2019polysemous,wu2020interpretable,zeng2022keyword}. In particular, \citet{su2021modeling} propose a dynamic intent graph (GRAPH4DIV) to balance content and intent in a document. \citet{song2019polysemous} utilize multiple transformers to extract visual features. \citet{wu2020interpretable} design an inactive word loss to expand the semantic concepts to represent various video contents. VMIG~\cite{zeng2022keyword} embeds each image and text query into multiple vectors via multiple instance learning. Although these methods succeed in boosting the semantic diversity of the retrieval results, they perform unsatisfactorily in guaranteeing semantic relevance and mining images of rare semantics.
\begin{table}
	\centering
	\caption{A qualitative comparison between CoLT and major existing methods from three dimensions: feature projection, retrieval result generation and performance.}
	\resizebox{0.48\textwidth}{!}{
	\begin{tabular}{c|ccc}
		\toprule
		Method & Projection & Generation & Performance\cr
		\midrule
		CLIP~\cite{radford2021learning} &One-to-one&  top-$k$ & Low diversity\\
		MMR~\cite{renders2017nle}   &One-to-one&  Re-ranking & Low diversity\\
		UMONS~\cite{seddati2017umons}  &One-to-one&Clustering& Low relevance\\
		VMIG~\cite{zeng2022keyword}  &Multi-vector& top-$k$ & Medium relevance \& diversity\\
		CoLT~(ours)  &SCL & TTC & High relevance \& diversity\\
		\bottomrule
	\end{tabular}
	}
	\label{tab-diff}
\end{table}

\subsection{Differences between Our Method and Existing Works}
To expound the differences between CoLT and typical existing methods, in Tab.~\ref{tab-diff} we present a qualitative comparison from three dimensions: how are images and queries projected? how are the final retrieval results generated? and how is the performance in terms of both relevance and diversity?
As presented in Tab.~\ref{tab-diff}, recent pre-trained cross-modal encoders (\textit{e.g.} CLIP~\cite{radford2021learning}) cannot model semantic diversity well due to the limitations of the one-to-one projection. Two typical post-processing based methods MMR and UMONS are poor at either modeling diversity~\cite{renders2017nle} due to the lack of an accurate diversity measurement mechanism or guaranteeing relevance due to clustering irrelevant features together. The recently proposed VMIG suffers from the robustness issue due to the uncertainty of multi-vector projection and the rare semantics handling issue caused by the top-$k$ search algorithm, which leads to undesirable performance. Our method CoLT is the only retrieval method that achieves both high semantic relevance and rich semantic diversity thanks to the proposed \emph{semantics-aware contrastive learning} (SCL) and powerful \emph{transformer-based token classifier} (TTC). Experiments and visualization studies demonstrate the advantages of our method.

\section{Methodology}

\subsection{Overview}
Given a text query $Q$ and an image dataset $\mathcal{D}$, our aim is to generate a retrieval list $\mathcal{R}$ that consists of $K$ images of high semantic relevance and diversity. Fig.~\ref{fig:pipeline} shows the architecture of our method CoLT, which is composed of six components: a \emph{fixed feature encoder} $f$ that takes $Q$ and $\mathcal{D}$ as input to generate initial query feature $h_q$ and visual features $\{ h_v^i \}$, \textit{i.e.,} \{$h_q, \{h_v^i\}\} = f(Q,\mathcal{D})$, a \emph{visual feature re-encoder} $g$ that re-encodes the visual features with the help of the \emph{semantics-aware contrastive learning} (SCL) module,
and the \emph{transformer-based token classifier} (TTC) $\phi$ that takes the query feature $h_q$ and the re-encoded image features $\hat{h}_v^i$ as an input token sequence to subsume each token into a suitable semantic category according to their representations. The TTC module consists of two sub-modules: the token classification transformer that is composed of $L$ transformer encoder layers, and a fully-connected layer as the classifier. Finally, a \emph{post-processing} module is adopted to select typical images from these categories as the final results. During training, a \emph{token-wise data augmentation} module is used to make full use of the training data, which is employed between the SCL module and the TTC module.

\subsection{Semantics-aware Contrastive Learning}
\label{sec:scl}
\begin{figure}[t]
    \centering
    \includegraphics[width=0.8\columnwidth]{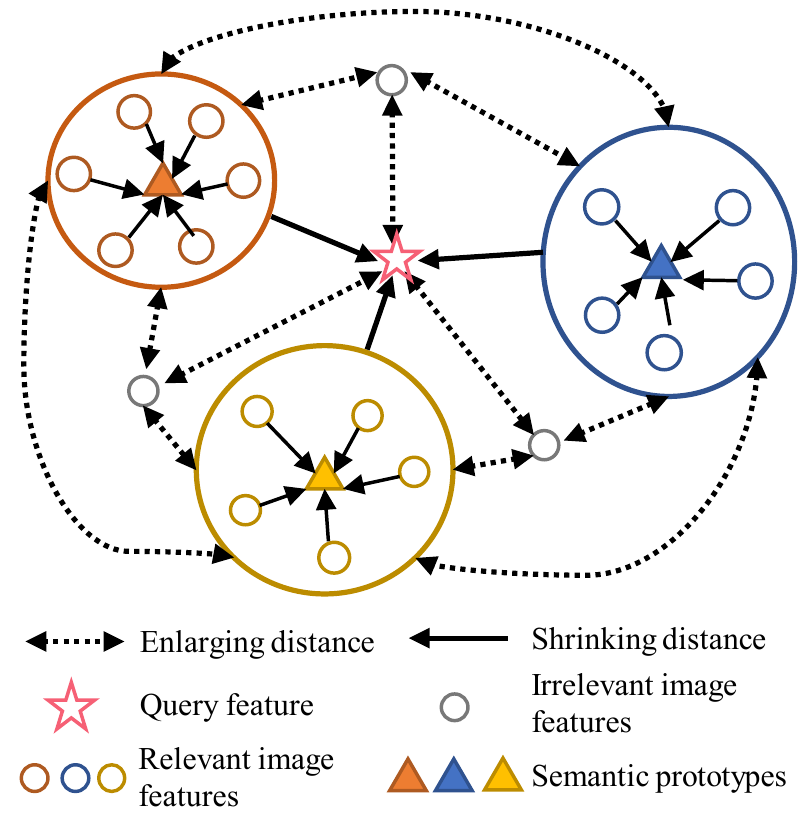}
    \caption{Illustration of semantics-aware contrastive learning. Different colors indicate various categories.}
    \label{fig:ccl}
\end{figure}
In CoLT, we first use a fixed pre-trained feature encoder $f$ to extract preliminary high-quality and robust visual features and query feature. Nevertheless, as mentioned above, these one-to-one projected features are not distinguishable enough to support effective diverse retrieval. Ergo, a visual feature re-encoder $g$, which is implemented by a multi-layer perception and powered by a novel semantics-aware contrastive learning is used to refine the image features to promote semantic diversity. In particular, for each visual feature $h_v^i$, we re-encode its representation as follows:
\begin{equation}
    \label{eq:revisor}
    \hat{h}_v^i=h_v^i + \beta g(h_v^i),
\end{equation}
where $\beta$ is a hyper-parameter used to control the learned re-encoded feature $g(h_v^j)$. In what follows, we introduce the proposed semantics-aware contrastive loss in detail.

As shown in Fig.~\ref{fig:ccl}, the goal of semantics-aware contrastive learning (SCL) is:  (1) Enlarging the distance between query feature and irrelevant features; (2) Enlarging the distance between relevant image features and irrelevant image features; (3) Enlarging the distance among image features of different semantics, which makes these features more distinguishable while benefiting diversity; (4) Shrinking the distance between the query feature and relevant image features, which can improve accuracy as (1) and (2); (5) Shrinking the distance among image features of similar semantics. In SCL, we use semantic category prototypes stored in a bank $\mathcal{B}$ to efficiently compute (3) and (5), which can avoid inputting a large batch size. As a result, each image query will be projected to a position with suitable distance between the query feature, its matched semantic prototypes, unmatched semantic prototypes, and irrelevant image features.

Here we discuss the implementation of the proposed semantics-aware contrastive learning. In SCL, the positive pairs include (1) relevant image-query feature pairs and (2) relevant image-category prototype pairs, while the negative pairs are (3) irrelevant image-query feature pairs and (4) irrelevant image-category prototype pairs. For a query $h_q$ with a set of relevant image features $\{\hat{h}_v^{r,i}\}$ and a set of irrelevant image features $\{\hat{h}_v^{ir,i}\}$. Let $\mathcal{B}(i)$ denotes the $i-$th semantic category prototype stored in the bank and $G(\cdot)$ is a function that maps the image features to the corresponding indices of the matched semantic category prototypes, the loss of SCL can be formulated as follows:
\begin{equation}
\label{eq:ccl}
\mathcal{L}_{scl} = - log\frac{
\overbrace{\Sigma_i exp(h_q \cdot \hat{h}_v^{r,i}/ \tau)}^{(1)} +
\overbrace{\Sigma_i exp(\mathcal{B}(G(\hat{h}_v^{r,i})) \cdot \hat{h}_v^{r,i}/ \tau)}^{(2)}
}
{\underbrace{\Sigma_i exp(h_q \cdot \hat{h}_v^{ir,i}/ \tau)}_{(3)} + \underbrace{\Sigma_{i,j} exp(\mathcal{B}(j) \cdot \hat{h}_v^{ir,i}/ \tau)}_{(4)}+ (1) + (2)},
\end{equation}
where $\tau$ is a hyper-parameter used to control the temperature.

The category prototypes stored in the bank $\mathcal{B}$ play an important role in the proposed SCL. Therefore, they need to be initialized and updated during training to obtain accurate and latest representations. Ergo, we use the fine-grained textual description features extracted by the fixed feature encoder to initialize the bank. As for update, exponential moving average (EMA)~\cite{klinker2011exponential,zhang2022hierarchical,ge2020self,zhu2020inflated} is utilized to update the category prototypes:
\begin{equation}
\label{eq:ema}
    \mathcal{B}(G(\hat{h}_v^{r,i})) = \alpha \mathcal{B}(G(\hat{h}_v^{r,i})) + (1-\alpha) \hat{h}_v^{r,i},
\end{equation}
where $\alpha$ is the momentum coefficient used to update the bank.

\subsection{Transformer-based Token Classification}
\label{sec:tct}
After obtaining a set of representation features, the next problem is how to generate the final result of high relevance and diversity. To this end, a powerful \emph{transformer-based token classifier} (TTC) is developed to do feature fusion and token classification. Specifically, we treat each feature as a token, and concatenate the query feature $h_{q}$ and $N$ image features $\{\hat{h}_{v}^i\}_{i=1}^{N}$ to form the input token sequence, \textit{i.e.,} $\mathcal{I}=[h_{q},\{\hat{h}_{v}^i\}_{i=1}^{N}]$. Here, to avoid irrelevant tokens, only $N$ image features semantically most similar to the query feature are used. We sort the image features with respect to their cosine similarity with the query feature, and generate the corresponding ground truth $\mathcal{Y}=\{y_i\}_{i=1}^{N+1}$ in terms of their fine-grained semantic annotations. It is worth mentioning that in TTC, all the irrelevant image features and the query feature are labeled with special indexes to further distinguish them.
As shown in Fig.~\ref{fig:pipeline}, $L$ transformer encoder layers~\cite{vaswani2017attention} powered by multi-head self-attention and feed forward layers are used to fuse these tokens. Subsequently, a fully connected layer is stacked as a classifier to do prediction. Formally, the predictions of TTC can be written as follows:
\begin{equation}
\label{eq:tct}
\{p_i\}_{i=1}^{N+1} = \phi(\mathcal{I}),
\end{equation}
where $p_i$ is the predicted distribution probability of the $i-$th token. Cross entropy loss is served as the classification loss to train TTC:
\begin{equation}
\label{eq:tct}
\mathcal{L}_{cls} = - \Sigma_i y_i log(p_i).
\end{equation}
After classifying each image token into an appropriate category, a post-processing algorithm $t$ is applied to generate the final retrieval list. That is, selecting $X$ images with the highest similarity to the query feature from each semantic category:
\begin{equation}
\label{eq:post-processing}
\mathcal{R} = t(\{p_i\}_{i=1}^{N+1},X).
\end{equation}
Finally, after selecting images from $\lfloor k/X \rfloor$ semantic categories, a retrieval list $\mathcal{R}$ of length $k$ is obtained.
\subsection{Token-wise Data Augmentation}\label{sec:da}
Due to the lack of specific fine-grained annotations, directly training our model on $\mathcal{D}$ is prone to over-fitting. Therefore, to better exploit the potential of the transformer-based token classification module, we employ token-wise data augmentation to enrich the input tokens $\mathcal{I}$. In particular, four different kernel operations are introduced:

\textbf{Query perturbation:} We perturb the query feature $h_q$ as follows:  MIXUP~\cite{zhang2017mixup} the query feature with a relevant image feature $\hat{h}_v^{r,i}$ with a probability of $p_q$. Formally, let $\lambda \sim Beta(1.0, 1.0) $, we generate the perturbed query feature as follows:
\begin{equation}
\label{eq:perturb_q}
h_q = max(\lambda, 1.0-\lambda) h_q + min(\lambda, 1.0-\lambda) \hat{h}_v^{r,i}.
\end{equation}

\textbf{Image perturbation:} We perturb the image feature $\hat{h}_v^{r,i}$ as follows: MIXUP the image feature with a relevant query feature $h_q$ with a probability of $p_v$. By sampling $\lambda$ from $Beta(1.0,1.0)$, we have
\begin{equation}
\label{eq:perturb_v}
\hat{h}_v^{r,i} = max(\lambda, 1.0-\lambda) \hat{h}_v^{r,i} + min(\lambda, 1.0-\lambda) \hat{h}_q.
\end{equation}

\textbf{Deletion:} We delete an image feature with a probability of $p_d$.

\textbf{Copy:} We copy an image feature with a probability of $p_c$.

Among the 4 operations, query perturbation and image perturbation directly augment the features without modifying the semantics-aware representations, which is beneficial to the robustness of the model while the operations of deletion and copy can enhance the model's ability of distinguishing rare and similar tokens, respectively. Following the experience in \cite{wei2019eda}, we perform data augmentation to the input tokens $\mathcal{I}$ in such a manner: sampling each data augmentation operation in the following order: (1) query perturbation; (2) deletion; (3) copy; (4) image perturbation, then individually performing the selected operation on each token.

\begin{algorithm}[t]
\caption{The training of CoLT.}
\begin{algorithmic}[1]
\State {\bf Input:}
Fixed feature encoder $f$, visual feature re-encoder $g$, transformer-based token classifier $\phi$, query $Q$, and image dataset $\mathcal{D}$
\State $h_q, \{h_v^i\} = f(Q,\mathcal{D})$
\State initialize $\mathcal{B}$ with fine-grained description
\While{$g$ is not convergenced}
    \State $\hat{h}_v^i = h_v^i + \beta g(h_v^i)$
    \State $\hat{h}_v^{r,i}, \hat{h}_v^{ir,i} \sim g(h_v^i)$
    \State Compute $\mathcal{L}_{scl}$ via Eq.~(\ref{eq:ccl})
    \State Optimize $g$ according to $\mathcal{L}_{scl}$
    \State Update $\mathcal{B}$ via Eq.~(\ref{eq:ema})
\EndWhile
\While{$\phi$ is not convergenced}
    \State $\mathcal{I}=[h_{q},\{\hat{h}_{v}^i\}_{i=1}^{N}]$
    \State Perform data augmentation to $\mathcal{I}$ according to Sec.~\ref{sec:da}
    \State Obtain the final $\mathcal{I}$ according to Sec.~\ref{sec:tct}
    \State $\{p_i\}_{i=1}^{N+1} = \phi(\mathcal{I})$
    \State Compute $\mathcal{L}_{cls}$ via Eq.~(\ref{eq:tct})
    \State Optimize $\phi$ according to $\mathcal{L}_{cls}$
\EndWhile
\State \Return $g$ and $\phi$
\end{algorithmic}
\label{alg:cct}
\end{algorithm}

\begin{algorithm}[t]
\caption{The evaluation of CoLT.}
\begin{algorithmic}[1]
\State {\bf Input:}
Fixed feature encoder $f$, visual feature re-encoder $g$, transformer-based token classifier $\phi$, query $Q$, image dataset $\mathcal{D}$, and post-processing algorithm $t$ with its hyper-parameter $X$
\State $h_q, \{h_v^i\} = f(Q,\mathcal{D})$
\State $\hat{h}_v^i = h_v^i + \beta g(h_v^i)$
\State $\mathcal{I}=[h_{q},\{\hat{h}_{v}^i\}_{i=1}^{N}]$
\State $\{p_i\}_{i=1}^{N+1} = \phi(\mathcal{I})$
\State $\mathcal{R} = t(\{p_i\}_{i=1}^{N+1},X)$
\State \Return $\mathcal{R}$
\end{algorithmic}
\label{alg:cct:eval}
\end{algorithm}

\subsection{Training and Evaluation Algorithms}
The training procedure of CoLT is presented in Alg.~\ref{alg:cct}, which can be divided into three steps. First, the initial query feature and image features are extracted by $f$ (L2). Then, we train the visual feature re-encoder by the proposed semantics-aware contrastive learning (L3-L9) to re-encode the preliminary features to semantics-aware ones. Finally, we take the query feature and the re-encoded image features as input to train the transformer-based token classifier with the help of token-wise data augmentation (L10-L16).

The evaluation procedure is given in Alg.~\ref{alg:cct:eval}, which is like this: we first generate the initial features (L2), then re-encode the image features (L3). Subsequently, take these features as tokens to generate the predicted distribution probabilities (L4-L5). Finally, using the post-processing algorithm $t$ to generate the final retrieval list $\mathcal{R}$.

\section{Performance Evaluation}
\subsection{Research Questions}
In this section, we evaluate the proposed method by conducting extensive experiments to answer the following research questions:\setlist{nolistsep}
\begin{itemize}
\item[\textbf{RQ1:}] How does CoLT perform in comparison with the state-of-the-art cross-modal image retrieval models in terms of both relevance and diversity?
\item[\textbf{RQ2:}] Can the proposed semantics-aware contrastive learning and the transformer-based token classifier effectively boost relevance and diversity?
\item[\textbf{RQ3:}] How do different components/parameters contribute to the effectiveness of CoLT?
\end{itemize}

\subsection{Datasets and Metrics}
Here we briefly summarize the datasets and metrics used in our paper. More details can be referred to~\cite{ionescu2021benchmarking}.
Two datasets are used in our paper:

\textbf{Div400:} Div400\footnote{\url{http://multimediaeval.org/mediaeval2014/diverseimages2014}} is collected by the MediaEval Workshop~\cite{inoescu2014div400}. It contains $396$ queries with $43,418$ images. All queries are mainly related to tourist locations and the average length of queries is $3.7$ words. On average, the ground truth of a query covers $11.8$ semantic categories of images in the dataset. Each image has a coarse-grained textual description (\textit{e.g.} ``Big Ben'') and a fine-grained one (\textit{e.g.} ``partial view'').

\textbf{Div150Cred:} Div150Cred\footnote{\url{http://campus.pub.ro/lab7/bionescu/Div150Cred.html}} is derived from the competition dataset for diverse social image retrieval in 2014 \cite{ionescu2015div150cred}. It has a total of $153$ queries with $45,375$ images. The ground truth of a query averagely covers $22.6$ semantic categories of images in the dataset.

Three metrics are used to evaluate the performance, including \emph{precision} ($P$) for measuring semantic relevance, \emph{cluster recall} ($CR$) for measuring semantic diversity, and the $F1$ \emph{score} of $P$ and $CR$ to measure the overall balanced performance. Specifically, we calculate the evaluation metrics of the top-$k$ results, where $k$ is set to 10 and 20 by following \cite{zeng2022keyword}. In the rest of this paper, we use P@k, CR@k, and F1@k to denote the $P$, $CR$, and $F1$ value of the top-$k$ results, respectively. Higher P@k indicates better relevance, and higher CR@k means richer semantic diversity.

\subsection{Implementation Details}
CoLT is implemented in PyTorch-1.10. All experiments are conducted on 4 NVIDIA 3090 GPUs with 24GB memory. The model is trained using the Adam~\cite{kingma2014adam} optimizer with a learning rate of $10^{-5}$ for the visual feature re-encoder $g$ and $10^{-4}$ for the transformer-base token classifier $\phi$. The batch size is set to 32. $\tau$, $\alpha$, $\beta$, and $\epsilon$ are set to small values: $\tau$ = $0.2$, $\alpha$ = $0.01$, $\beta$ = $0.02$, and $\epsilon$=$0.01$ by following~\cite{zhao2021recursive,klinker2011exponential,ge2020self,zhu2020inflated}. $X$, $N$, and $L$ are set to 1, 200, and 8 through ablation study. The probabilities used for data augmentation are set by following~\cite{wei2019eda}. In particular, we have $p_q$ = $0.5$, $p_v$ = $0.2$, $p_d$ = $0.2$ and $p_c$ = $0.2$. All different semantic categories (or simply semantics) in each dataset are stored as prototypes. As a result, we store 629 prototypes for Div400 dataset while 725 for Div150Cred dataset.

\begin{table}[t]
\centering
\caption{Performance comparison with the state-of-the-art methods on Div400.
P@k and CR@k are evaluation metrics for relevance and diversity, respectively.
F1@k evaluates the overall performance regarding both relevance and diversity.
}
\label{tab:sota:div400}
\newsavebox{\overalll}
\begin{lrbox}{\overalll}
\resizebox{0.48\textwidth}{!}{
\begin{tabular}{c|c|c|c|c|c|c}
\hline
Method      & \begin{tabular}[c]{@{}c@{}}P@10\end{tabular} & \begin{tabular}[c]{@{}c@{}}P@20\end{tabular} &
\begin{tabular}[c]{@{}c@{}}CR@10\end{tabular} & \begin{tabular}[c]{@{}c@{}}CR@20\end{tabular} &
\begin{tabular}[c]{@{}c@{}}F1@10\end{tabular} & \begin{tabular}[c]{@{}c@{}}F1@20\end{tabular}
\\
\hline
IMRAM~\cite{chen2020imram} & 79.22\% & 76.98\%  & 28.93\% & 33.79\% &42.38\% &46.96\% \\
FCA-Net~\cite{han2021fine} & 79.98\% & 78.42\% & 29.91\% & 34.90\% & 43.54\% & 48.30\%\\
CLIP~\cite{radford2021learning} & \textbf{90.17\%} & \textbf{87.92\%} & 35.68\% & 52.97\% & 51.10\% & 65.60\% \\
\hline
MMR~\cite{renders2017nle} & \underline{86.79\%} & 84.52\% & 36.96\% & 53.88\% & 51.85\% & 65.81\%\\
UMONS~\cite{seddati2017umons} & 79.28\% & 73.37\% & \underline{44.71\%} & \underline{63.24\%} & \underline{57.17}\% & 67.93\%\\
\hline
DESA~\cite{qin2020diversifying} & 76.50\% & 73.82\% & 39.47\% & 52.82\% &52.07\% &61.58\%\\
Su \textit{et al.}~\cite{su2021modeling}  & 77.55\% & 74.50\% & 35.66\% & 45.77\% &48.85\% &56.70\%\\
VMIG~\cite{zeng2022keyword} & 82.31\% & 83.01\% & 40.37\% & 59.46\% &54.17\% &\underline{69.28\%} \\
\hline
CoLT~(ours)& 84.45\% & \underline{85.48\%} & \textbf{47.06\%} & \textbf{64.16\%} &\textbf{60.44}\% &\textbf{73.30}\% \\
\hline
\end{tabular}
}
\end{lrbox}
\scalebox{0.99}{\usebox{\overalll}}
\end{table}

\subsection{Comparing with SOTA Methods (RQ1)}
To demonstrate the effectiveness of our method CoLT, we compare it with several state-of-the-art approaches, including three typical cross-modal image retrieval methods: IMRAM~\cite{chen2020imram}, FCA-Net~\cite{han2021fine} and CLIP~\cite{radford2021learning}, two post-processing-based diverse retrieval methods: MMR~\cite{renders2017nle} and UMONS~\cite{seddati2017umons}, and three learning-based diverse retrieval approaches: DESA~\cite{qin2020diversifying}, GRAPH4DIV~\cite{su2021modeling} and VMIG~\cite{zeng2022keyword}. Since MMR, UMONS and VMIG require a feature encoder, for fairness, we use the same feature encoder CLIP~\cite{radford2021learning} to implement them and our method CoLT. For MMR and UMONS, we use grid search to obtain their best results. Generally, our results are higher than those in the original papers thanks to the strong feature encoder. For example, the P@20, CR@20, and F1@20 values of VMIG on the DIV400 dataset are lifted from 78.27\%, 59.01\% and 67.29\% to 83.01\%, 59.46\% and 69.28\%.
Experimental results on Div400 and Div150Cred are given in Tab.~\ref{tab:sota:div400} and Tab.~\ref{tab:sota:div150}, respectively. Here, the best values are bolded while the second-best results are underlined.

From Tab.~\ref{tab:sota:div400} and Tab.~\ref{tab:sota:div150}, we can see that 1) typical cross-modal image retrieval methods including large-scale pre-trained encoder CLIP perform well in relevance-based retrieval but cannot do diverse retrieval well. For example, although CLIP achieves the best relevance performance, it is inferior to the others in diversity score. 2) Post-processing-based approaches can only moderately trade-off accuracy and diversity. For example, as can be seen in Tab.~\ref{tab:sota:div400}, the diversity improvement achieved by MMR is very limited (CR@10 increases from 35.68\% to 36.96\% on Div400). As for UMONS, its accuracy score is greatly degraded (P@10 decreases from 90.17\% to 79.28\% on Div400) though it obtains a relatively large diversity improvement. As a result, their $F1$ scores are not satisfactory. 3) Recently proposed learning-based methods achieve balanced relevance and diversity scores. For instance, VMIG outperforms most existing methods in CR@10 and CR@20, and performs better than UMONS in relevance score. However, its relevance and diversity are both limited due to the weaknesses of the multi-vector projection. 4) Our method CoLT obtains the best diversity score, high precision, and obviously the highest overall $F1$ score on both Div400 and Div150Cred. In particular, CoLT outperforms CLIP and VMIG by significant margins, i.e., 7.70\% and 4.02\% of F1@20 on Div400, respectively. This indicates that CoLT is able to get retrieval results of both high relevance and rich semantic diversity. Besides, we present a variant of CoLT that outperforms CLIP on both relevance and diversity, we will discuss the details in Sec.~\ref{sec:ab}.

\begin{table}[t]
\centering
\caption{Performance comparison with the state-of-the-art methods on Div150Cred.}
\label{tab:sota:div150}
\resizebox{0.48\textwidth}{!}{
\begin{tabular}{c|c|c|c|c|c|c}
\hline
Method      & \begin{tabular}[c]{@{}c@{}}P@10\end{tabular} & \begin{tabular}[c]{@{}c@{}}P@20\end{tabular} &
\begin{tabular}[c]{@{}c@{}}CR@10\end{tabular} & \begin{tabular}[c]{@{}c@{}}CR@20\end{tabular} &
\begin{tabular}[c]{@{}c@{}}F1@10\end{tabular} & \begin{tabular}[c]{@{}c@{}}F1@20\end{tabular}
\\
\hline
CLIP~\cite{radford2021learning} & \textbf{96.02\%} & \textbf{95.04\%} & 23.48\% & 35.32\% & 37.73\% & 51.51\% \\
\hline
MMR~\cite{renders2017nle} & \underline{95.37\%} & 94.23\% & 23.50\% & 35.49\% & 37.71\% & 51.56\%\\
UMONS~\cite{seddati2017umons} & 77.40\% & 84.15\% & \underline{26.69}\% & \textbf{40.10\%} & \underline{39.69}\% & \underline{54.32\%}\\
\hline
VMIG~\cite{zeng2022keyword} & 90.81\% & 89.96\% & 23.83\% & 37.97\% &37.75\% &53.40\% \\
\hline
CoLT~(ours)& 93.41\% & \underline{94.39\%} & \textbf{27.53\%} & \underline{39.30\%} &\textbf{42.52}\% &\textbf{55.49}\% \\
\hline
\end{tabular}
}
\scalebox{0.99}{\usebox{\overalll}}
\end{table}

\begin{figure*}[t]
    \centering
    \includegraphics[width=1.4\columnwidth]{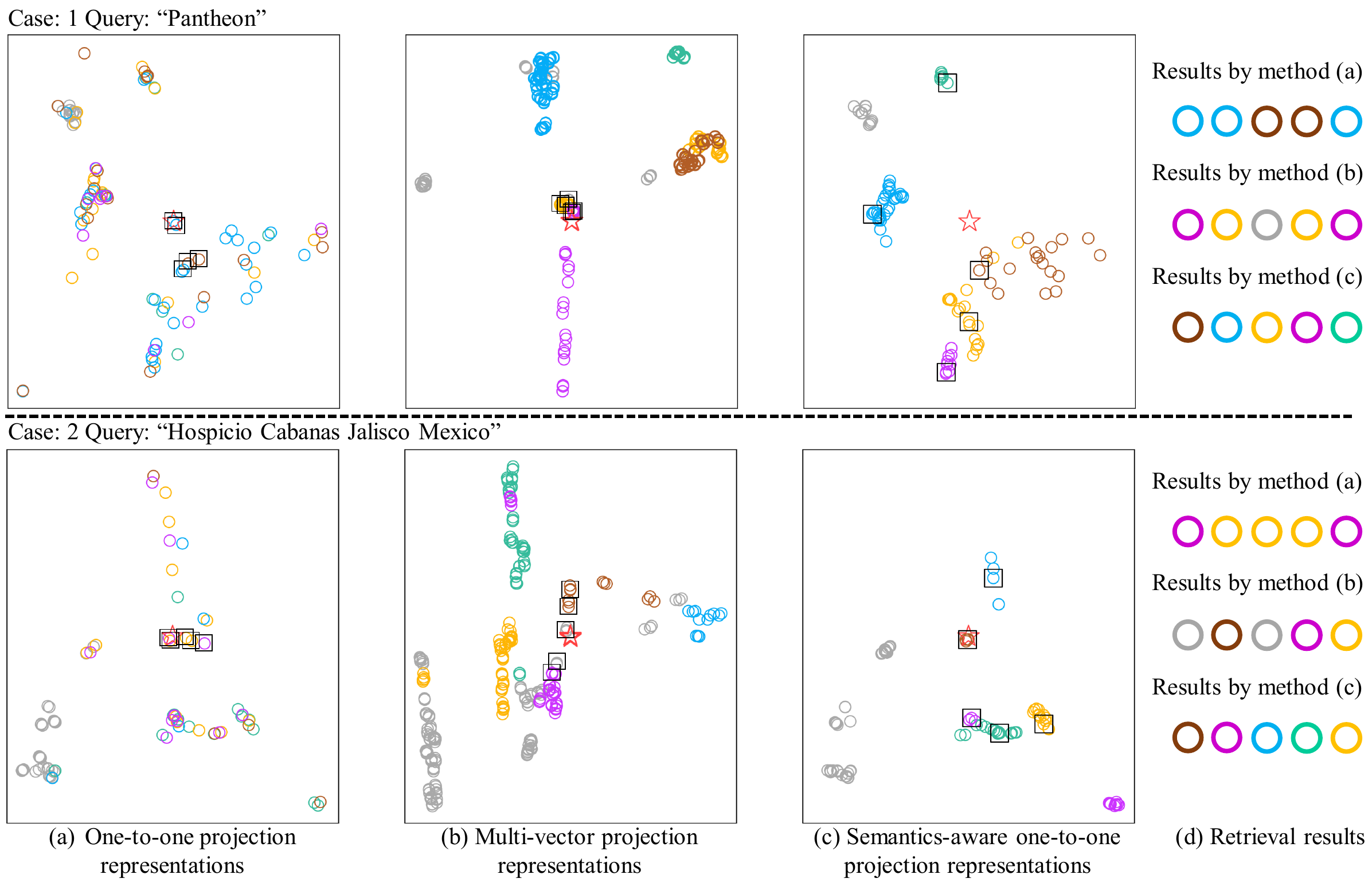}
    \caption{Visualization comparison of different representations. (a) One-to-one projection (OOP) representations generated by the cross-modal encoder $f$. (b) Multi-vector projection (MVP) representations. (c) Semantics-aware one-to-one projection (SA-OOP) representations re-encoded by $g$. Retrieved images are marked by black square. (d) The final retrieval results generated by different methods.}
    \label{fig:visualization-1}
\end{figure*}
\begin{figure*}[t]
    \centering
    \includegraphics[width=1.4\columnwidth]{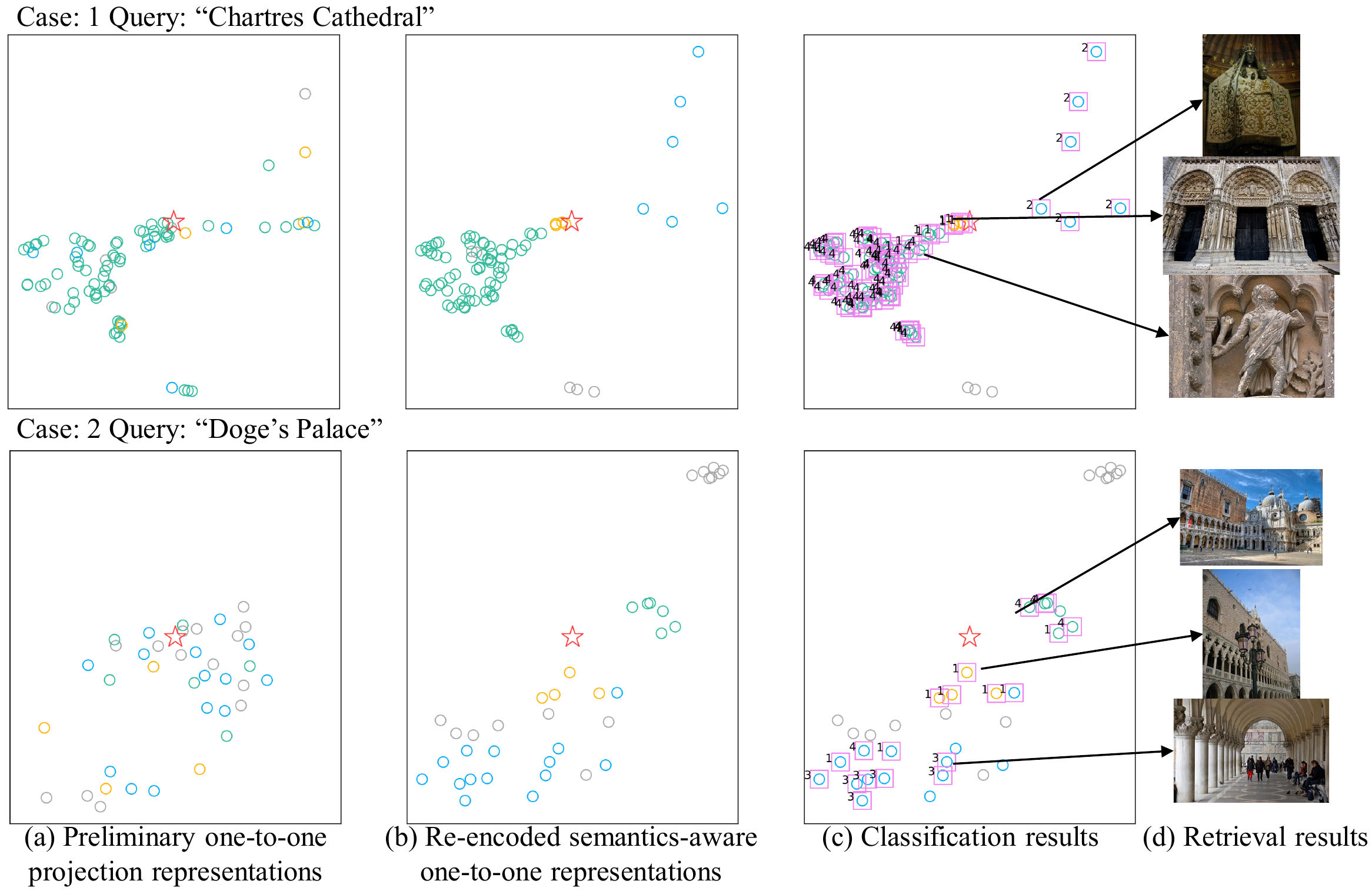}
    \caption{The visualization of CoLT results. (a) One-to-one projection (OOP) representations generated by a cross-modal encoder $f$. (b) Semantics-aware one-to-one projection (SA-OOP) representations generated by the re-encoder $g$. (c) Classification results of TTC $\phi$. To make the figures clear, irrelevant images are not marked. The numbers around the boxes are the category ID predicted by TTC. (d) The final retrieval results obtained by our post-processing algorithm.}
    \label{fig:visualization-2}
\end{figure*}

\subsection{Visualization (RQ2)}
To better demonstrate the advantages of the proposed techniques and study how they lift the performance, we visualize some cases in the test set of the DIV400 dataset via UMAP~\cite{mcinnes2018umap}. The visualization comparison between our semantics-aware representation and existing methods' representations are illustrated in Fig.~\ref{fig:visualization-1}.

From Fig.~\ref{fig:visualization-1}(a), we can see that the preliminary OOP representations extracted by CLIP can distinguish some irrelevant images. However, its weaknesses are also evident: (1) The query is closer to some image features of common semantics (the blue and brown points in the 1st case); (2) Images of various semantics are mixed. As a result, such representations are not suitable for mining diversity. Then, let us pay attention to multi-vector projection (MVP). As can be seen from Fig.~\ref{fig:visualization-1}(b), each image and query are projected into multiple points to enrich diversity. However, on the one hand, some outliers are mistakenly projected into the neighborhood of the query feature to represent diversity (a grey point in the 1st case while two in the 2nd case). On the other hand, some image features of rare semantics are projected into remote regions (the green points in the 1st case) where the top-$k$ algorithm cannot reach. Thus, as shown in Fig.~\ref{fig:visualization-1}(d), some irrelevant images are selected while some images of rare semantics are not retrieved. Finally, we check the representations of our SCL and the images retrieved by TTC. From Fig.~\ref{fig:visualization-1}(c) we can see that (1) the representations of images of the same semantics are clustered and much more distinguishable compared with the typical OOP representations in Fig.~\ref{fig:visualization-1}(a); (2) Some irrelevant images are also pushed away. For instance, in the 2nd case, some grey points are pushed to the left-bottom corner. This demonstrates the advantages and effectiveness of our SCL. Then, TTC and a post-processing are employed to classify and select images from each category, including rare semantic categories like green points in the 1st case, to form the final results.

We also visualize the classification results of TTC to further demonstrate its effect. The visualization is shown in Fig.~\ref{fig:visualization-2}, from which we can see that (1) TTC is able to distinguish different semantics and irrelevant images. Taking the 1st case for example, the yellow points are classified into the 1st category, the majority of the green points are subsumed into the 4th category, and blue points to the 2nd category. Irrelevant images are also correctly classified. This demonstrates the effectiveness of the proposed TTC. (2) The classification performance of TTC can be further improved. For example, as can be seen in the 2nd case, TTC mistakenly classifies one of the green points into the 1st category. In summary, the power of TTC is demonstrated well via visualization.

\subsection{Ablation Study (RQ3)}\label{sec:ab}
Here we conduct ablation study on Div400 to demonstrate the contributions of different modules and the effect of some parameters in our method. The metrics P@20, CR@20 and F1@20 are used. Results are presented in from Tab.~\ref{tab:ab} to Tab.~\ref{tab:L}.
\begin{table}[t]
\caption{Ablation study of CoLT on Div400.}
\centering
\scalebox{0.8}{
\begin{tabular}{c|c|ccc}
\hline
ID & Variant & P@20 &CR@20 &F1@20\\
\cline{1-5}
 \hline
1&without SCL\&TTC &87.92\% &52.97\% &65.60\% \\
2&SCL + TTC &85.48\% &64.16\% &73.30\% \\
\hline
3&without SCL &84.26\% &62.94\% &72.06\% \\
4& UMONS &73.37\% &63.24\% &67.93\% \\
5&SCL + DBSCAN&75.94\% &63.90\% &69.40\% \\
6&SCL + top-$k$  &89.06\% &89.93\% &67.79\%\\
\hline
7&without DA &86.34\% &62.19\% &72.30\% \\
\hline
8&unfixed $f$ &76.52\% &51.25\% &61.38\% \\
\hline
\end{tabular}}
\label{tab:ab}
\end{table}
\\
\begin{table}[t]
\caption{Effect of pair construction in SCL.}
\centering
\scalebox{0.75}{
\begin{tabular}{c|ccc}
\hline
Variant & P@20 &CR@20 &F1@20\\
\cline{1-4}
 \hline
All 4 pairs &85.48\% &\textbf{64.16}\% &\textbf{73.30}\% \\
Without pair (2) &85.04\% &63.78\% &72.89\% \\
Without pair (4) &\textbf{85.72}\% &62.17\% &72.07\% \\
\hline
\end{tabular}}
\label{tab:sclterm}
\end{table}
\noindent \textbf{Overall performance improvement.}
As shown in the 1st row and 2nd row in Tab.~\ref{tab:ab}, our method significantly boosts the diversity score and $F1$ score from 52.97\% to 64.16\% and 65.60\% and 73.30\%, respectively, with only a slight decrease in relevance score. This supports the superiority of our method.\\
\noindent \textbf{Effect of SCL.}
Here we check the effect of the proposed SCL. Specifically, we first design a variant that removes SCL and directly applies TTC. Obviously, as can be seen in the 2nd row and the 3rd row of Tab.~\ref{tab:ab}, all metrics including relevance, diversity, and $F1$ score are degraded. Besides, we also design a variant that combines SCL with the idea of UMONS to generate the final retrieval results via DBSCAN. Comparing the results of the 4th row and the 5th row, the performance with SCL is better than that of the original UMONS. The reason lies in that SCL is able to make the image features more distinguishable, and such representations are more suitable for existing post-processing schemes.

Then, we check the effect of the constructed pairs in SCL. As mentioned in Sec.~\ref{sec:scl}, SCL uses 4 kinds of pairs. Among these pairs, (1) and (3) are common in contrastive learning~\cite{radford2021learning} to align images and queries while (2) and (4) play important roles in distinguishing images of various semantics. Ergo, we remove (2) and (4) separately to examine their influence. As can be seen in Tab.~\ref{tab:sclterm}, without pair (2) and pair (4), diversity score and F1 score are degraded. On the contrary, their influence on relevance score is minor. This justifies the effectiveness of SCL --- making the representations much more distinguishable for promoting diversity. \\
\noindent \textbf{Effect of TTC.}
To check the effect of the proposed transformer-based token classifier, we design two variants that replace TTC by DSCAN (the 5th row of Tab.~\ref{tab:ab}) or top-$k$ (the 6th row of Tab.~\ref{tab:ab}) to generate the retrieval results. Obviously, such variants are inferior to our method (the 2nd row of Tab.~\ref{tab:ab}). This demonstrates the advantage of TTC.\\
\noindent \textbf{Effect of token-wise data augmentation.} Here we implement a variant that removes the token-wise data augmentation module. Results of this variant are given in the 7th row of Tab.~\ref{tab:ab}. Evidently, the resulting performance is inferior to ours (the 2nd row of Tab.~\ref{tab:ab}).\\
\noindent \textbf{Why fix the cross-modal feature encoder?} In CoLT, we fix the cross-modal feature encoder $f$ to better maintain the pre-trained knowledge. To support this design, we implement a variant that finetunes the feature encoder $f$. Experimental results are given in the 8th row of Tab.~\ref{tab:ab}. Obviously, all metrics including relevance, diversity and $F1$ are significantly degraded, comparing with ours (the 2nd row of Tab.~\ref{tab:ab}). Possibly, finetuning too many parameters is prone to over-fitting.
\\
\begin{table}[t]
\caption{Performance when using different feature encoders.}
\centering
\scalebox{0.8}{
\begin{tabular}{c|ccc}
\hline
Variant & P@20 &CR@20 &F1@20\\
\cline{1-4}
 \hline
ViT+BERT &87.75\% &52.39\% &65.60\% \\
+CoLT &85.48\% &64.16\% &73.30\% \\
\hline
R50+BERT &87.01\% &52.10\% &65.17\% \\
+CoLT &85.62\% &60.83\% &72.94\% \\
\hline
GroupViT & 83.68\% &52.02\% &64.21\% \\
+CoLT &82.62\% &70.91\% &70.91\% \\
\hline
\end{tabular}}
\label{tab:ffe}
\end{table}
\noindent\textbf{Can CoLT support various feature encoders?} As mentioned above, CoLT is general, i.e., it can work with various cross-modal encoders to do diverse image retrieval. To verify this point, we try three different encoder configurations, including ViT~\cite{dosovitskiy2020image} and BERT~\cite{devlin2019bert} developed by \cite{radford2021learning}, R50~\cite{he2016deep} and BERT~\cite{devlin2019bert} implemented by \cite{radford2021learning}, and the encoders proposed in GroupViT~\cite{xu2022groupvit}. The experimental results are given in Tab.~\ref{tab:ffe}, from which we can see that (1) all these pre-trained cross-modal encoders are good at relevance-based retrieval but perform poorly in terms of CR@20; (2) After applying our method CoLT, the diversity score is significantly boosted, with only slight decrease in precision. As a result, superior F1 score is achieved. This validates that CoLT can work well with various feature encoders to boost performance.\\
\begin{table}[t]
\caption{Performance \emph{vs}. the number of images selected from each category.}
\centering
\scalebox{0.8}{
\begin{tabular}{c|ccc}
\hline
$X$ & P@20 &CR@20 &F1@20\\
\cline{1-4}
 \hline
1 &85.48\% &64.16\% &73.30\% \\
2 &88.09\% &58.54\% &70.34\% \\
3 &88.45\% &57.28\% &69.54\% \\
\hline
\end{tabular}}
\label{tab:numbercluster}
\end{table}
\textbf{Can CoLT flexibly balance accuracy and diversity?}
As mentioned above, we can flexibly trade-off the relevance and diversity of the retrieval results without modifying network parameters. This is achieved by controlling the hyper-parameter $X$. As described in Sec.~\ref{sec:tct}, the post-processing algorithm will select $X$ images from each semantic category to form a retrieval list $\mathcal{R}$ of length $k$. Thus, a smaller $X$ will select fewer images from each category but can include more different categories (estimated by $\lfloor k/X \rfloor$), which will benefit the diversity of the retrieval list $\mathcal{R}$ but may hurt the relevance since classification accuracy on rare semantic categories is poor. On the contrary, a larger $X$, i.e., selecting more images from each category of common semantics will benefit the accuracy but limit the semantic diversity since fewer categories are exploited. We present the experimental results of how $X$ impacts performance in Tab.~\ref{tab:numbercluster}. We can see that the best diversity is achieved when $X=1$ while the best accuracy is obtained when $X$=3. This indicates that CoLT can meet various retrieval settings, which demonstrates the flexibility of our approach. In this paper, we set $X$=1 by default to obtain the best diversity and $F1$ score.\\
\begin{table}[t]
\caption{Time cost comparison. We report the result of one-time retrieval on a 3090 GPU.}
\centering

\scalebox{0.8}{
\begin{tabular}{c|ccccc}
\hline
~&CLIP & MMR &UMONS&VMIG&CoLT (Ours)\\
\cline{1-6}
Time~(ms) &18.06&24.10&22.65&86.77&30.23\\
\hline
\end{tabular}}
\label{tab:timecp}
\end{table}
\begin{table}[t]
\caption{Time cost comparison among major components. We report the result of one time retrieval on a 3090 GPU. $f$ and $g$ are tested in a parallel manner.}
\centering

\scalebox{0.8}{
\begin{tabular}{c|ccc}
\hline
~ & $f$ &$g$&$\phi$\\
\cline{1-4}
Time~(ms) &18.06 &0.37 &11.80\\
\hline
\end{tabular}}
\label{tab:time}
\end{table}
\textbf{Time cost.} We first compare the time cost of our method with that of various SOTA methods. The experimental results are given in Tab.~\ref{tab:timecp}. On the one hand, our method CoLT is 2.87$\times$ faster than the state-of-the-art learning-based method VMIG. On the other hand, our method consumes moderately more time than the post-processing-based methods. For example, CoLT takes 6.23ms more than MMR. This justifies the efficiency of our method.

Then, we further check the time cost of each major module in CoLT: the fixed feature encoder $f$, the visual feature re-encoder $g$, and TTC $\phi$. The experimental results are given in Tab.~\ref{tab:time}. We can see that $g$ and $\phi$ incur much less time than the feature encoder $f$. The reason lies in that $g$ is a simple multi-layer perceptron while $\phi$ consists of multiple transformer encoder layers that can run in parallel. It is worthy of mentioning that the image features generated by $f$ and $g$ can be cached offline in application. Hence, the main cost is from TTC $\phi$, which is very limited (11.80ms according to Tab.~\ref{tab:time}). This also verifies the efficiency of our method.\\
\begin{table}[t]
\caption{The effect of parameter $L$ in TTC.}
\centering
\scalebox{0.8}{
\begin{tabular}{c|cccc}
\hline
$L$ & P@20 &CR@20 &F1@20 &Time (ms)\\
\cline{1-4}
 \hline
6 &85.82\% &61.34\% &71.54\% &8.93\\
8 &85.48\% &64.16\% &73.30\% &11.80\\
10 &85.29\% &61.67\% &71.58\% &18.56\\
\hline
\end{tabular}}
\label{tab:L}
\end{table}
\noindent \textbf{Effect of the parameter $L$.}
Here we study the effect of the number of transformer layers $L$. On the one hand, a larger $L$ may result in over-fitting at a higher probability due to the limited training data. On the other hand, a smaller $L$ cannot fully exploit the potential of TTC. Therefore, we conduct a grid search to determine the value of $L$. As can be seen in Tab.~\ref{tab:L}, the best performance is achieved when $L$=8.
\\
\noindent \textbf{Effect of the parameter $N$.}
Here we check how the number of images $N$ fed to the transformer-based token classifier  $\phi$ impacts the performance. Intuitively, a large $N$ will include images with more semantics. On the other hand, a large $N$ will introduce more irrelevant images that may make token classification more difficult. On the contrary, a small $N$ includes less irrelevant images but also fewer semantics. Therefore, both small $N$ and large $N$ are not appropriate for TTC. We conduct grid search to determine $N$ on two datasets. Based on our results, we set $N$ to 200 for the DIV400 dataset because the F1@20 scores of $N=150$ and $N=250$ are 72.10\% and 72.13\%, which is inferior to that of $N=200$ where F1@20 is 73.30\%. While on the DIV150Cred dataset, the best performance is achieved when $N=200$ (an $F1$ of 55.49\%) and $250$ (an $F1$ of 55.32\%). Ergo, we set this hyper-parameter to 200.\\

\section{Conclusion}
In this paper, we address keyword-based diverse image retrieval and propose a new method called Semantics-aware Classification Transformer (CoLT) to do this task. Different from existing works, CoLT first extracts highly representative images and query features via semantics-aware contrastive learning, then a transformer-based token classifier is employed to fuse these features and subsume them into their appropriate categories. Finally, a post-processing algorithm is applied to flexibly selecting images from each category to form the retrieval results. The advantages of CoLT are four-fold: \textit{high semantic relevance}, \textit{high semantic diversity}, \textit{general} and \textit{easy-to-use}, and \textit{easy-to-control}. Extensive experiments on two datasets Div400 and Div150Cred demonstrate the superiority of our method.
\begin{acks}
Minyi Zhao was supported in part by the 2022 Tencent Rhino-Bird Research Elite Training Program. Shuigeng Zhou was supported by National Key R\&D Program of China under grant No.~2021YFC3340302.
\end{acks}

\newpage
\bibliographystyle{ACM-Reference-Format}
\balance
\bibliography{main}
\clearpage
\end{document}